\documentclass[a4paper,showpacs,notitlepage,nobibnotes,endfloats,preprint,footinbib,final,onecolumn,12pt]{revtex4}
\usepackage{amssymb}

\usepackage{amsmath}

\newtheorem{theorem}{Theorem}
\newtheorem{acknowledgement}[theorem]{Acknowledgement}

\input{tcilatex}

\begin{document}

\title{Study of the decoherence of a double quantum dot charge qubit via the
Redfield equation}
\author{Zeng-Zhao Li, Xiao-Yin Pan, Xian-Ting Liang}
\email{xtliang@ustc.edu;Tel:+86-574-87600783; Fax:+86-574-87600744}
\affiliation{Department of Physics and Institute of Modern Physics, Ningbo University,
Ningbo 315211, China}

\begin{abstract}
By using the Redfield form of the master equation, we investigate the
decoherence times of a double quantum dot charge qubit (DQDCQ) in three
different cases, namely when it is coupled to (I) the piezoelectric coupling
phonon bath (PCPB), (II) the deformation coupling phonon bath (DCPB), and
(III) the Ohmic bath. It is found that our results for case (I) and (II) are
in the same magnitude with those obtained via the exact path integral
methods, while for case (III), the decoherence time is in well agreement
with the experimental value.

\end{abstract}

\maketitle

\section{Introduction}

The double quantum dot (DQD) charge qubit [1-5] is one of the qubits that
are considered to be promising candidates for the realization of the
building blocks of quantum information processing. Its two low-energy states
are denoted as the local states $\left| 0\right\rangle $ and $\left|
1\right\rangle $, which could be controlled via external voltage sources.
There exist some effective schemes to prepare the initial states and readout
the final states of the qubit [6]. It is also known that this kind of qubit
is coupled to its environment inevitably. Therefore, it is believed that the
decoherence might be the central impediment for the qubit to be taken as the
cell of quantum computer. Hence, finding out the primary origin or
dominating mechanism of decoherence for the qubit is a basic task to
overcome the difficulty.

The recently experimental realization of the coherent manipulation of
electronic states in a double-dot system [7, 8], which was implemented in a
GaAs/AlGaAs heterostructure containing a two-dimensional electron gas,
stimulated a lot of theoretical interests. Various methods have been tried
to study the system. For instance, through density matrix simulation,
Fujisawa et al. [9] tried to explain its transportation property. The
Born-Markov-type electron-phonon decoherence at large times due to
spontaneous phonon emission of the quantum dot charge qubits was
investigated by Fedichkin et al. [10]. In 2005, Vorojtsov et al. [11]
studied the decoherence of the DQD charge qubit by employing the Born-Markov
approximation. Based on a unitary transformation, Wu et al. [12]
investigated the decoherence in terms of a perturbative treatment. Thorwart
et al. [13] investigated the decoherence of the DQD charge qubit in a longer
time with a numerical exact iterative quasiadiabatic propagator path
integral (QUAPI) method [14], while Liang [15] used an iterative tensor
multiplication (ITM) method [14] derived from the QUAPI to study decoherence
of a DQD charge qubit in both the piezoelectric coupling phonon bath (PCPB)
and the deformation coupling phonon bath (DCPB). They found that the
decoherence times of the qubit are shorter than the reported experimental
ones when the qubit in PCPB and DCPB.

In this paper, we investigate the decoherence times of the (DQD) charge
qubit coupled to PCPB, DCPB and the Ohmic bath with another method, through
solving the master equations. The model Hamiltonian and the spectral
functions for the different baths are introduced in section II. We then
develop the Redfield form of the master equation and obtain the decoherence
times in section III. Conclusions are given in the last section.

%It is shown that the decoherence time obtained when the
%environment is PCPB or DCPB  has the same order of magnitude  as
%those obtained by the exact path integral method.  However, this
%method is much easier than the numerical path integral one.
%Moreover, we study the decoherence time of the qubit in the Ohmic
%bath, our results are in well agreement with the experimental
%ones. \cite{r7,r8}.

\section{QDQ charge qubit model}

%In this section, we shall introduce the model Hamiltonian for the
%double quantum dot and then develop the Redfield equation theory
%leading to the explicit expression for the evolution of the
%density matrix elements.\\

The DQD charge qubit consists of left and right dots connected through an
interdot tunneling barrier. Due to the Coulomb blockade, at most one excess
electron is allowed to occupy the left and right dot, which defines two
basis vectors $\left\vert 0\right\rangle $ and $\left\vert 1\right\rangle $.
The energy difference between these two states can be controlled by the
source-drain voltage. Neglecting the higher order tunneling between leads
and the dots, the effective Hamiltonian in the manipulation process reads
[12, 13] 
\begin{equation*}
H_{eff}=H_{S}+H_{B}+H_{SB},
\end{equation*}%
where $H_{S}$ is the Hamiltonian of the QDQ charge qubit, $H_{B}$ is the
Hamiltonian for the phonon bath, and $H_{SB}$ describes the electron-phonon
interaction. More explicitly, we have 
\begin{equation*}
H_{S}=\hbar T_{c}\sigma _{x},
\end{equation*}%
\begin{equation*}
H_{B}=\hbar \sum_{q}\omega _{q}b_{q}^{\dagger }b_{q},
\end{equation*}%
\begin{equation}
H_{SB}=\hbar \sigma _{z}\sum_{q}(M_{q}b_{q}^{\dagger }+M_{q}^{\ast }b_{q}).
\label{1}
\end{equation}%
Here, $T_{c}$ is the interdot tunneling, $\sigma _{x}\ $and $\sigma _{z}$\
are the Pauli matrix, $b_{q}^{\dagger }$ ($b_{q}$) are the creation
(annihilation) operators of phonons, $\hbar \omega _{q}$\ is the energy of
the phonons, and $M_{q}=C_{q}/\sqrt{2m_{q}\omega _{q}\hbar }$, where $C_{q}$
are the classical coupling constants of the qubit-phonons system. Having
written above, we now introduce the spectral density which fully describes
the effects of the phonon bath [16, 17] 
\begin{equation}
J(\omega )=\sum_{q}\left\vert M_{q}\right\vert ^{2}\delta \left( \omega
-\omega _{q}\right) .  \label{2}
\end{equation}%
According to Ref. [12], the spectral density of PCPB is 
\begin{equation}
J^{pz}(\omega )=g_{pz}\omega \left[ 1-\frac{\omega _{d}}{\omega }\sin \left( 
\frac{\omega }{\omega _{d}}\right) \right] e^{-\omega ^{2}/2\omega _{l}^{2}},
\label{3}
\end{equation}%
where $\omega _{d}=s/d$ and $\omega _{l}=s/l$, $d$\ denotes the
center-to-center distance of two dots, $l$\ the dot size, $s$ the sound
velocity in the crystal, and 
\begin{equation}
g_{pz}=\frac{M}{\pi ^{2}\varrho s^{3}}\left( \frac{6}{35}+\frac{1}{x}\frac{8%
}{35}\right) .  \label{4}
\end{equation}%
As usual, $M$\ is the piezoconstant, $\varrho $\ is the density of the
crystal,\ and $x$\ is the rate of transverse to the longitudinal of sound
velocity in the crystal (see for example Refs. [12] and [13]). As in the
GaAs crystal $s\approx 5\times 10^{3}$ $($m/s$)$. With the parameters of
GaAs [18], Wu et al. [12] estimated that $g_{pz}\approx 0.035$ $\left( \text{%
ps}\right) ^{-2}$. The spectral density of DCPB is also obtained. It is 
\begin{equation}
J^{df}(\omega )=g_{df}\omega ^{3}\left[ 1-\frac{\omega _{d}}{\omega }\sin
\left( \frac{\omega }{\omega _{d}}\right) \right] e^{-\omega ^{2}/2\omega
_{l}^{2}},  \label{5}
\end{equation}%
with 
\begin{equation*}
g_{df}=\frac{\Xi ^{2}}{8\pi ^{2}\varrho s^{5}},
\end{equation*}%
where $\Xi $\ is the deformation potential. In the same paper, Wu et al.
also propose a value $g_{df}\approx 0.029$ $($ps$)^{-2}$. With the help of
the definite spectral density functions of the baths, one can investigate
the dynamics and then the decoherence of the open qubit.

We also consider explicitly the case in which the behavior of the original
spectral function $J(\omega )$\ has a simple power-law form for $\omega \leq
\omega _{c}$: 
\begin{equation}
J(w)=\eta \omega ^{s}e^{-\omega /\omega _{c}},\text{ }\eta =const,  \label{6}
\end{equation}%
with dimensionless damping strength $\eta $ and a cutoff frequency $\omega
_{c}$. On general grounds, the linear low frequency behavior of $J(w)$ is
expected in basically all condensed-phase electron transfer (ET) reactions
[16], and the frequency $\omega _{c}$ then corresponds to some dominant bath
mode. For its analytic advantages, we use the Ohmic spectral density [17] by
setting $s=1$ in Eq. (6). We also study the case when $\eta =0.04$, $\omega
_{c}=0.05$ $($ps$)^{-1}$ [19].

\section{Decoherence times}

In this section, we first review the Redfield form of the master equation.
In an open system, the relevant system is characterized by the reduced
density-matrix (RDM) $\rho $ which is defined as a trace over all bath
variables of the full density matrix $\rho _{T}$ 
\begin{equation}
\rho =tr_{B}\left( \rho _{T}\right) .  \label{7}
\end{equation}%
The equation for the time evolution of $\rho $ can be consequently obtained
from the Liouville equation for $\rho _{T}$ 
\begin{equation}
\frac{\partial \rho (t)}{\partial t}=\frac{i}{\hslash }tr_{B}[\rho
_{T}(t),H].  \label{8}
\end{equation}%
The Redfield form of the master equation for the RDM is obtained from Eq.
(8) by implementing a series of perturbative approximations [20, 21],
namely, the system-bath coupling is treated perturbatively up to the second
order. The bath is assumed to remain in equilibrium, and then Markov
approximation gets involved. In the eigenstate representation of the system 
\begin{equation}
H_{s}\left| \mu \right\rangle =E_{s}\left| \mu \right\rangle ,  \label{9}
\end{equation}%
the Redfield equation for the RDM reads [22] 
\begin{equation}
\frac{\partial \rho _{\mu \nu }(t)}{\partial t}=-i\omega _{\mu \nu }\rho
_{\mu \nu }(t)+\sum_{\kappa \lambda }R_{\mu \nu \kappa \lambda }\rho
_{\kappa \lambda }(t),  \label{10}
\end{equation}%
where $\omega _{\mu \nu }=(E_{\mu }-E_{\nu })/\hbar $ and $R_{\mu \nu \kappa
\lambda }$ is the relaxation or Redfield tensor. The first term on the
right-hand side of Eq. (10) describes the isolated system evolution, while
the second one represents its interaction with the dissipative environment.
The Redfield tensor describing the system relaxation can be expressed as 
\begin{equation}
R_{\mu \nu \kappa \lambda }=\Gamma _{\lambda \nu \mu \kappa }^{+}+\Gamma
_{\lambda \nu \mu \kappa }^{-}-\delta _{\upsilon \lambda }\sum_{\alpha
}\Gamma _{\mu \alpha \alpha \kappa }^{+}-\delta _{\mu \kappa }\sum_{\alpha
}\Gamma _{\lambda \alpha \alpha \upsilon }^{-},  \label{11}
\end{equation}%
with 
\begin{equation}
\Gamma _{\lambda \nu \mu \kappa }^{+}=\frac{1}{\hslash ^{2}}\int_{0}^{\infty
}dt\langle \left\langle \lambda \right| H_{SB}(t)\left| \upsilon
\right\rangle \left\langle \mu \right| H_{SB}\left| \kappa \right\rangle
\rangle _{B}e^{-i\omega _{\mu \kappa }t},  \label{12}
\end{equation}%
\begin{equation}
\Gamma _{\lambda \nu \mu \kappa }^{-}=\frac{1}{\hslash ^{2}}\int_{0}^{\infty
}dt\langle \left\langle \lambda \right| H_{SB}\left| \upsilon \right\rangle
\left\langle \mu \right| H_{SB}(t)\left| \kappa \right\rangle \rangle
_{B}e^{-i\omega _{\lambda \upsilon }t},  \label{13}
\end{equation}%
\begin{equation}
H_{SB}(t)=e^{iH_{B}t/\hslash }H_{SB}e^{-iH_{B}t/\hslash },  \label{14}
\end{equation}%
where $\langle ...\rangle _{B}$ denotes the thermal average over the bath.

For the Hamiltonian defined above, the Redfield tensor components by using
coherent states can explicitly be expressed as 
\begin{eqnarray}
\Gamma _{\lambda \nu \mu \kappa }^{+} &=&\frac{1}{2}\left\langle \lambda
\right\vert \sigma _{Z}\left\vert \upsilon \right\rangle \left\langle \mu
\right\vert \sigma _{Z}\left\vert \kappa \right\rangle J(\omega _{\kappa \mu
})(1+n(\omega _{\kappa \mu })),\text{ \thinspace\ if }\omega _{\kappa \mu
}>0,  \notag \\
\Gamma _{\lambda \nu \mu \kappa }^{+} &=&\frac{1}{2}\left\langle \lambda
\right\vert \sigma _{Z}\left\vert \upsilon \right\rangle \left\langle \mu
\right\vert \sigma _{Z}\left\vert \kappa \right\rangle J(\omega _{\kappa \mu
})n(\omega _{\mu \kappa }),\text{ \ \ \ \ \ \ \thinspace\ if }\omega _{\mu
\kappa }>0,  \label{15}
\end{eqnarray}%
\begin{eqnarray}
\Gamma _{\lambda \nu \mu \kappa }^{-} &=&\frac{1}{2}\left\langle \lambda
\right\vert \sigma _{Z}\left\vert \upsilon \right\rangle \left\langle \mu
\right\vert \sigma _{Z}\left\vert \kappa \right\rangle J(\omega _{\lambda
\upsilon })(1+n(\omega _{\lambda \upsilon })),\text{ \thinspace\ if }\omega
_{\lambda \upsilon }>0,  \notag \\
\Gamma _{\lambda \nu \mu \kappa }^{-} &=&\frac{1}{2}\left\langle \lambda
\right\vert \sigma _{Z}\left\vert \upsilon \right\rangle \left\langle \mu
\right\vert \sigma _{Z}\left\vert \kappa \right\rangle J(\omega _{\upsilon
\lambda })n(\omega _{\upsilon \lambda }),\text{ \ \ \ \ \ \ \thinspace\ if }%
\omega _{\upsilon \lambda }>0,  \label{16}
\end{eqnarray}%
where $\omega _{\mu }=E_{\mu }/\hslash $ are the system eigenfrequencies, $%
n(\omega )=1/(e^{\hslash \omega /\kappa T}-1)$ is the distribution function
for bosons, $\kappa $ is the Boltzmann constant, $T$ is the temperature, and 
$J(\omega )$ is the bath spectral function.

To measure effects of the decoherence, one can use the entropy, the first
entropy, and many other measures, such as maximal deviation norm, etc.( for
example, see Refs. 23, 24, 25). However, the decoherence of an open quantum
system is essentially reflected through the decays of the off-diagonal
coherent terms of its RDM. In general, the decoherence is produced due to
the interaction of the quantum system with other system which has a large
number of degrees of freedom, such as the devices of the measurement or
environment. Here, we investigate the decoherence times via directly
describing the evolutions of the off-diagonal coherent terms instead of
using any measure of decoherence. In the following, we set the initial state
of the qubit to $\rho (0)=\frac{1}{2}(\left| 0\right\rangle +\left|
1\right\rangle )(\left\langle 0\right| +\left\langle 1\right| )$, which is a
pure state and it has the maximum coherent terms, and the initial state of
the environment is $\rho _{bath(0)}=\Pi _{k}e^{-\beta M_{k}}/Tr(e^{-\beta
M_{k}})$, where $M_{k}=\omega _{k}b_{k}^{\dagger }b_{k}$, $\beta =1/\kappa
T. $ According to Ref. [12], we set $\omega _{d}=0.02$ $($ps$)^{-1}$, $%
T_{c}=0.1\omega _{l}$ in the calculations.

We then shall use the Redfield equation to investigate the decoherence time
of the DQD charge qubit. The analytical expressions of the elements of the
RDM we obtained are 
\begin{equation}
\rho _{11}(t)=\frac{1+n(\omega _{21})}{1+2n(\omega _{21})}-\frac{e^{-2\chi t}%
}{2(1+2n(\omega _{21}))},  \label{17}
\end{equation}%
\begin{equation}
\rho _{22}(t)=\frac{n(\omega _{21})}{1+2n(\omega _{21})}+\frac{e^{-2\chi t}}{%
2(1+2n(\omega _{21}))},  \label{18}
\end{equation}%
\begin{eqnarray}
\rho _{12}(t) &=&\frac{(\chi +\sqrt{\chi ^{2}-\omega _{21}^{2}})e^{(-\chi +%
\sqrt{\chi ^{2}-\omega _{21}^{2}})t}-(\chi -\sqrt{\chi ^{2}-\omega _{21}^{2}}%
)e^{(-\chi -\sqrt{\chi ^{2}-\omega _{21}^{2}})t}}{4\sqrt{\chi ^{2}-\omega
_{21}^{2}}}  \notag \\
&&+i\omega _{21}\frac{e^{(-\chi +\sqrt{\chi ^{2}-\omega _{21}^{2}}%
)t}-e^{(-\chi -\sqrt{\chi ^{2}-\omega _{21}^{2}})t}}{4\sqrt{\chi ^{2}-\omega
_{21}^{2}}},
\end{eqnarray}%
\begin{eqnarray}
\rho _{21}(t) &=&\frac{(\chi +\sqrt{\chi ^{2}-\omega _{21}^{2}})e^{(-\chi +%
\sqrt{\chi ^{2}-\omega _{21}^{2}})t}-(\chi -\sqrt{\chi ^{2}-\omega _{21}^{2}}%
)e^{(-\chi -\sqrt{\chi ^{2}-\omega _{21}^{2}})t}}{4\sqrt{\chi ^{2}-\omega
_{21}^{2}}}  \notag \\
&&-i\omega _{21}\frac{e^{(-\chi +\sqrt{\chi ^{2}-\omega _{21}^{2}}%
)t}-e^{(-\chi -\sqrt{\chi ^{2}-\omega _{21}^{2}})t}}{4\sqrt{\chi ^{2}-\omega
_{21}^{2}}},
\end{eqnarray}%
where $\chi =\hslash J(\omega _{21})(1+2n(\omega _{21}))/2$. Note that by
substituting the parameters $\omega _{21}$, $\omega _{d},$ $\omega _{l},$ $%
\omega _{c}$, $\eta $ and $T$, in PCPB, DCPB and Ohmic bath into the above
expressions, we can readily plot the evolutions of the $\rho _{12}$ (or $%
\rho _{21}$).

The evolutions of the coherent elements of the RDM of the DQD charge qubit
in PCPB and DCPB with different $\omega _{l}$ are plotted in Figs. 1 and 2
respectively. And when the environment is modeled with Ohmic bath the
evolution of the coherent elements of the RDM is plotted in Fig. 3 as
damping strength $\eta =0.04,$ $0.08$ and $0.12$.%
\begin{eqnarray*}
&& \\
&&Fig.1,Fig.2,Fig.3. \\
&&
\end{eqnarray*}%
It is shown that the decoherence times of the qubit decreases as the $\omega
_{l}$ increases in PCPB and DCPB models. And the decoherence time of the
qubit decreases as the damping strength $\eta $ in the Ohmic bath model
increases. From Figs. 1-3, we see that the decoherence times of the DQD
charge qubit in PCPB and DCPB are much shorter than the experimentally
suggested result, which is in agreement with the results of Refs.[13, 15].
However, if we choose the environment as the Ohmic bath, the decoherence
time of the qubit agrees with the experimental result when we choose a
suitable coupling coefficient of the qubit to the environment. Moreover, we
also consider the evolutions of the coherent term at different temperatures
when the bath is modeled by PCPB, DCPB and Ohmic bath separately, which are
plotted in Figs. 4, 5. and 6, respectively.%
\begin{eqnarray*}
&& \\
&&Fig.4,Fig.5,Fig.6. \\
&&
\end{eqnarray*}%
It is shown that the decoherence times decreases as the temperature
increases no matter which bath is used to model the environment when $\omega
_{l}$\ (in PCPB and DCPB) and $\eta $ (in Ohmic bath) are fixed, this is
physically reasonable.

\section{Conclusions}

In summary, we have investigated the decoherence times of a double quantum
dot (DQD) charge qubit when it is coupled to different baths by using the
Redfield equation method. Our results show that the qubits have shorter
decoherence times than the experimental ones as the environment is modeled
by the acoustic phonon baths, which agrees with previous reports. Moreover,
when we use Ohmic bath model the environment with the coupling coefficient
of the qubit to the environment properly chosen, the decoherence time of the
qubit is in well agreement with the experimental result.

\begin{acknowledgement}
This project was sponsored by National Natural Science Foundation of China
(Grant No. 10675066) and K.C.Wong Magna Foundation in Ningbo University.
\end{acknowledgement}

\section{Figures Captions}

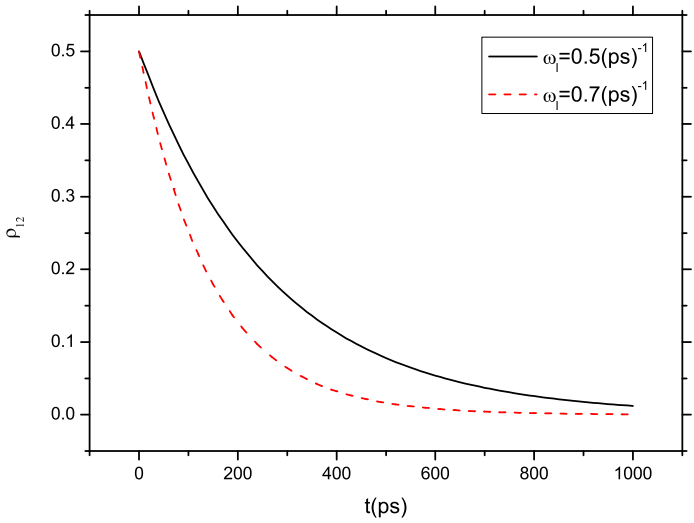
Fig. 1:\ $\rho _{12}(t)$ vs time $t$. The evolution of the off-diagonal
elements of the RDM for the DQD charge qubit in PCPB when $\omega _{l}=0.5$ $%
($ps$)^{-1}$ and $\omega _{l}=0.7$ $($ps$)^{-1}$. Here, $\omega _{d}=0.02$ $%
( $ps$)^{-1}$, $g_{pz}=0.035$ $($ps$)^{-2}$, and $T=30$ mK. The initial
state is described in the text.

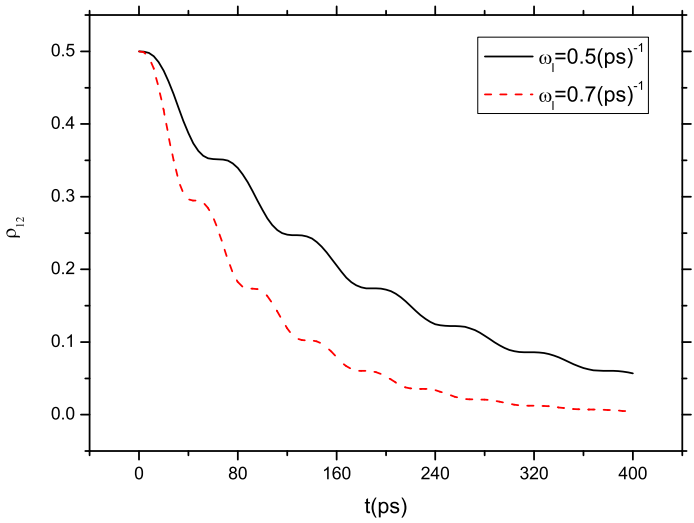
Fig. 2:\ $\rho _{12}(t)$ vs time $t$. The evolution of the off-diagonal
elements of the RDM for the DQD charge qubit in DCPB. Here, $\omega
_{d}=0.02 $ $($ps$)^{-1}$, $g_{df}=0.029$ $($ps$)^{-2}$. Other parameters
are the same as those in Fig. 1.

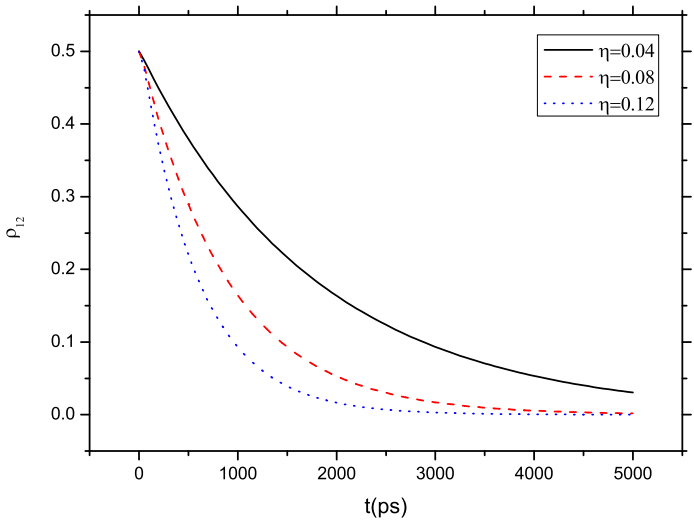
Fig. 3: $\rho _{12}(t)$ vs time $t$. The evolution of the off-diagonal
elements of the RDM for the DQD charge qubit in Ohmic case, with cut-off
frequency $\omega _{c}=0.05$ $($ps$)^{-1}$, $T=30$ mK and $\eta =0.04$, $%
0.08 $, $0.12$,\ respectively. The initial state is described in the text.

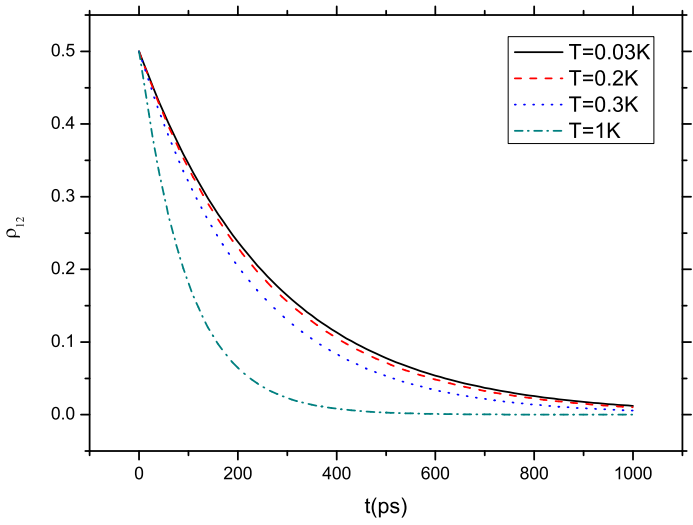
Fig. 4:\ $\rho _{12}(t)$ vs time $t$. The evolution of the off-diagonal
elements of the RDM for the DQD charge qubit in PCPB when $\omega _{l}=0.5$ $%
($ps$)^{-1}$, $T=30$ mK, $200$ mK, $300$ mK, $1$ K respectively. Other
parameters are the same as those in Fig. 1.

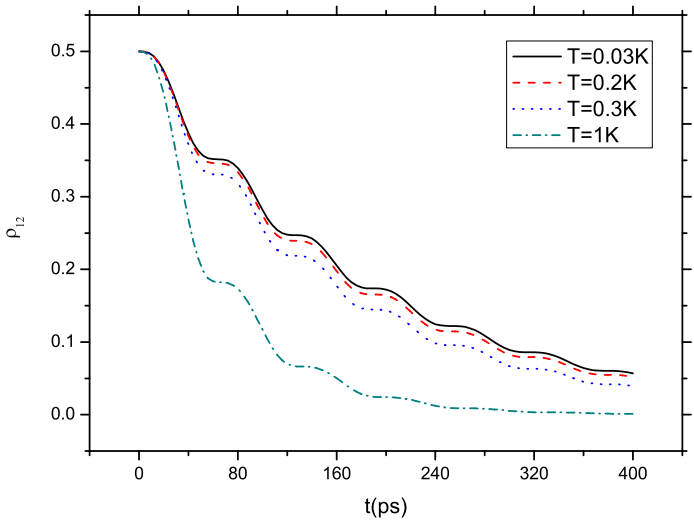
Fig. 5:\ $\rho _{12}(t)$ vs time $t$. The evolution of the off-diagonal
elements of RDM for the DQD charge qubit in DCPB when $\omega _{l}=0.5$ $($ps%
$)^{-1}$, $T=30$ mK, $200$ mK, $300$ mK, $1$ K respectively. Other
parameters are the same as those in Fig. 2.

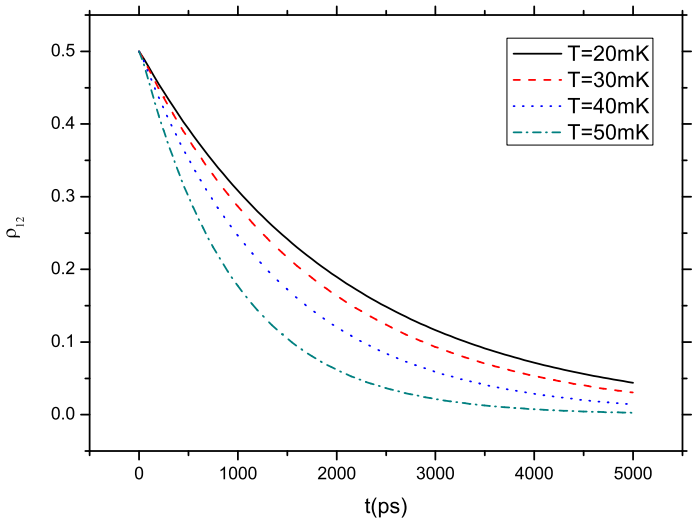
Fig. 6: $\rho _{12}(t)$ vs time $t$. The evolution of the off-diagonal
elements of the RDM for the DQD charge qubit in Ohmic case when $\eta =0.04$
and $T=20$ mK, $30$ mK, $40$ mK, $50$ mK,\ respectively. Other parameters
are the same as those in Fig. 3.

\end{document}